# The Hydrogen Atom of Bohr I:
# Own Life Time of an Excited Single Atom.

P. S. Kamenov

Many of the authors think that the theory of Bohr about the hydrogen atom is not adequate to reality. Bohr himself thinks (and writes) that, though his model of hydrogen describes some properties of this atom very well, it can not be accepted as realistic... Now it is considered to be **almost a shame** to think about this model as corresponding to reality. Some textbooks for students discuss this model of Bohr, but only as a curious example of an inadequate model which can lead to some true results[1]...

In this paper *I hope to show that this understanding has no foundations*. It stems from some mystic assumption about quantum physics [2]...For example, the complementarity principle postulated by Niels Bohr [3] in 1927, assumes that the wave-particle "duality" is a property of a single quantum system and therefore its two complementary aspects cannot be measured simultaneously; In some experiments (interference, diffraction) wave properties are manifested, some other experiments provide evidence for the corpuscular properties of quantum objects (impulse, energy). It is impossible to observe the two phenomena simultaneously and, moreover, they *cannot exist simultaneously* ... In the paper [4] it was shown that in the case of waves on the surface of a liquid the floating classical particles which pass through only one of the two opened slits are guided by the interfering surface waves in the same directions as predicted by quantum laws. This is a confirmation of de Broglie's ideas that *wave and particle exist simultaneously and that this coexistence is real* [5]. Most of the scientists think that the field of de Broglie is not real and they accept the statistical interpretation of Born [6]. I ask myself why the assumption that a wave-particle can not exist simultaneously is more real than de Broglie's assumption that they **always exist simultaneously** [5]?. De Broglie's waves in the hydrogen atoms are such that in the stationary state the mass of the electron (m), its velocity $v_n$ , and average radius $r_n$ are related with the principal quantum number (n= 1,2,3....) according to:

(1) $$mv_n r_n = n\frac{h}{2\pi} = n\hbar$$

and the field-particle (electron) is in a potential well which guides the electron only in orbit *n*, in which case the electron does not emit photons. The length of de Broglie's wave $\lambda_n$ exactly satisfies the condition

(2) $$2\pi r_n = \frac{nh}{mv_n} = n\lambda_n$$

The "wave-particle" electron is bound together with the "wave-particle" proton by electromagnetic forces. They interfere and remain in their potential well (position) forever, like classical particles on the surface of a liquid [4].The field of de Broglie is so real and this reality is so strong that the electromagnetic forces can not destroy this interference and the field-particle (electron) can not emit a photon-soliton [7,8]. In order to explain the decay of a stationary state it is necessary to assume some infinitely small "external perturbation" which would disturb the exact equalities (1) and (2) and (after some time of destructive interference) permit the transition to lower states. Only the ground state ($2\pi r_1 = \lambda_1$; n=1) can not be disturbed by a "infinitely small perturbation" because the field-particle can not be destroyed (n can not be smaller than 1). In this case only if the perturbation energy is **sufficiently** great, the electron can make a transition from the ground to the upper levels (absorption only, [9]). Excited by some energy, the electron can randomly occur at any distance ($r_{ni}$) *around* the exact radius of the stationary orbit ($2\pi r_{ni} \approx n\lambda_n$;). We can imagine that the "wave-particle" electron self-interferes as long as the minima of the wave coincide with the maxima of the precedent waves (this means that $|D(t)|^2 = 0$; D is the amplitude of the interfering electron waves). In this moment a transition occurs and energy is emitted. The greater the difference $|r_{ni} - r_n|$, the smaller the time necessary for destructive interference. If $|r_{ni} - r_n| \to 0$, the time for destructive interference would be very long [10]. *It seems that all particles (including photons [7,8]) must have a charge which I call "charge of de Broglie". This charge creates a monochromatic wave with the length of the wave of de Broglie* $\lambda = \hbar/\mathrm{mv}$. *As it is known, all other fields created by charged particles (when moving with constant velocity v) can be represented by monochromatic waves with this* $\lambda$.

**Return to Bohr's atom and the real field-particle of de Broglie.** In Fig.1 a schematic wave-particle in some excited state of the hydrogen atom is shown. The particle-wave electron moves from left to right (for example, n=2). In Fig.1 a) the velocity of the electron $v_n$ is such that $\lambda_n$ and $r_n$ correspond exactly to Bohr's conditions:

$$(3) \quad \lambda_n = \frac{h}{\mathrm{mv}_n}$$

Such a wave-particle electron returns from the left always with the same phase and reiterates its motion for an infinitely long time. If the velocity of the electron (v) is slightly different, the new $\lambda$ will also be slightly different (compared with $\lambda_n$):

$$(4) \quad \lambda = \frac{h}{\mathrm{mv}}$$

Such a particle-wave electron would arrive from the left (Fig. 1 b)) with a slightly different phase. With time this difference increases and we can calculate the moment when the sum of two amplitudes becomes zero (for the first time). In this moment the electron is not more in the potential well of the wave (like classical particles, [4], when $|D|^2 = 0$) and is allowed to change its position. This moment will be the time of life of this excited atom. The sum of the two amplitudes of de Broglie' field (D) can be written (like with classical particles, [4]):

(5) $$D = \sin\left(\frac{2\pi}{\lambda}(vt - r)\right) + \sin\left(\frac{2\pi}{\lambda}(vt)\right)$$

where r is the new radius which is only slightly different from $r_n$. The relation between $\lambda$, $\omega$ and v is:

(6) $$\lambda = \frac{2\pi v}{\omega}$$

One can substitute this in (5) and obtain:

(7) $$D = \sin\left(\omega t - \frac{\omega r}{v}\right) + \sin(\omega t)$$

Because of r/v=1/$\omega$, (7) becomes

(8) $$D = \sin(\omega t - 1) + \sin(\omega t)$$

which is the sum of de Broglie's amplitudes (D), expressed by the time and the frequency of a not exactly stationary state. From (3) and (4) one can find the small difference $\Delta\lambda$ and $\Delta\omega$:

(9) $$\frac{\Delta\lambda}{\lambda} = \frac{\lambda_n - \lambda}{\lambda} = \left(\frac{v}{v_n} - 1\right) = \frac{\omega}{\Delta\omega}$$

Taking into account that

(10) $$\frac{v}{v_n} = \sqrt[3]{\frac{\omega}{\omega_n}} = \sqrt[3]{\frac{\omega_n + \Delta\omega}{\omega_n}}$$

and from (9) we obtain ($\omega$):

(11) $$\omega = \left(\Delta\omega\left(\sqrt[3]{\frac{\omega_n + \Delta\omega}{\omega_n}} - 1\right)\right)$$

As usual, we must find the moment (t) when $|D|^2 = 0$ (the electron is not in the potential well of its wave):

(12) $$|D|^2 = |\sin(\omega t - 1) + \sin(\omega t)|^2 = 0$$

Hence

(13) $$\sin(\omega t - 1) = -\sin(\omega t)$$

or

$$\omega t - 1 = -\omega t ; \qquad t = \frac{1}{2\omega}$$

So, substituting $\omega$ from (11), we find the necessary "own lifetime" (t):

(14) $$t = \frac{1}{2\Delta\omega\left(\sqrt[3]{\frac{\omega}{\omega_n}} - 1\right)} = \frac{1}{2\Delta\omega\left(\sqrt[3]{\frac{\omega_n + \Delta\omega}{\omega_n}} - 1\right)}$$

As it is seen, when $\omega = \omega_n$ (or $\Delta\omega = 0$), the time is t→∞, as it should be for a stationary state. For $\Delta\omega << \omega_n$, the expression fo r the time (14) is symmetric (for positive and negative $\Delta\omega$). It is more convenient (for me) to transform eq.(14) in terms of energy:

(15) $$t = \frac{\hbar}{2\Delta E\left(\sqrt[3]{\frac{\Delta E}{E_n} + 1} - 1\right)}$$

where the energy can be measured in units eV and $\hbar$ [eV.s]. In this case the energy of the different excited states can be expressed through the Rydberg constant (R). Thus, the life time of each single excited hydrogen atom depends on the small energy difference ($\Delta E$) and the principal quantum number (n):

(16) $$t = \frac{\hbar}{2\Delta E \left( \sqrt[3]{1 + \frac{n^2 \Delta E}{R}} - 1 \right)}$$

In the case when $n^2 \Delta E \ll R$, the cubic root can be expanded in a series, and taking only two first terms of the expansion $(1 + n^2 (\Delta E)/3R ...)$ we obtain:

(17) $$t_f = \frac{3\hbar R}{2(\Delta E)^2 n^2}$$

Part of the results are shown in the Fig.2. These curves are different for different excited states (n). They could be compared with the normalized "own lifetimes" of nuclei ($t/\tau$ and $\Delta E / \Gamma$) [10]. Like in this work [10], one can find the mean life time ($\tau_n$) of an ensemble of excited hydrogen atoms.

(18) $$\tau_n = \frac{\hbar}{\Gamma_n} = n \sqrt{\frac{\hbar}{24 \pi R}}$$

For n=2, $\tau_n = 1.603 \times 10^{-9}$ s, which coincides exactly with the data from [11] (where $\tau_n = 1.60 \times 10^{-9}$ s).

***Conclusions.*** These results show that if one accepts de Broglie's assertion about the reality of the unitary wave-particle, the excited *single hydrogen atom decays (after excitation) in an exactly predictable moment (17); this moment depends upon the small energy difference ($\Delta E$) between the stationary energy level and the individual excitation energy. The mean life time ($\tau_n$) is a characteristic only of a statistical ensemble of excited atoms.* The complementarity principle is not necessary, and the statistical interpretation of quantum mechanics should be applied only to ensembles of atoms. The experimental proof of these results is possible with the help of resonant Mossbauer experiments in the time domain, as it was mentioned in [10].

This work was supported in part by the Bulgarian National Foundation for Scientific Investigations (No 534/1995).

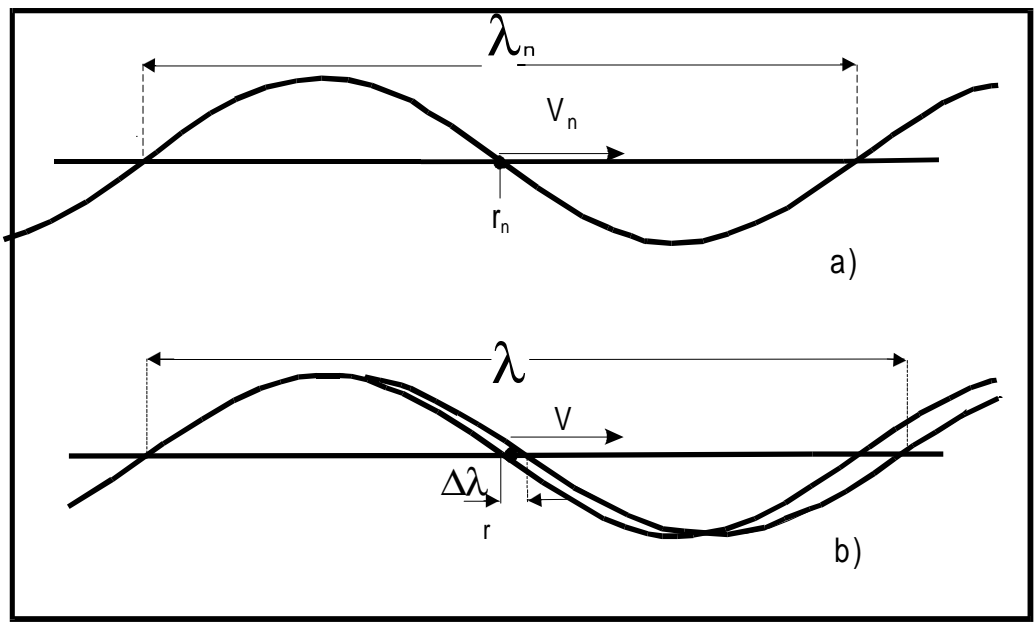

Fig.1. A scheme of one of the first hydrogen excited states. Wave-particle electron and its interference; a) true stationary state; b) almost stationary state.

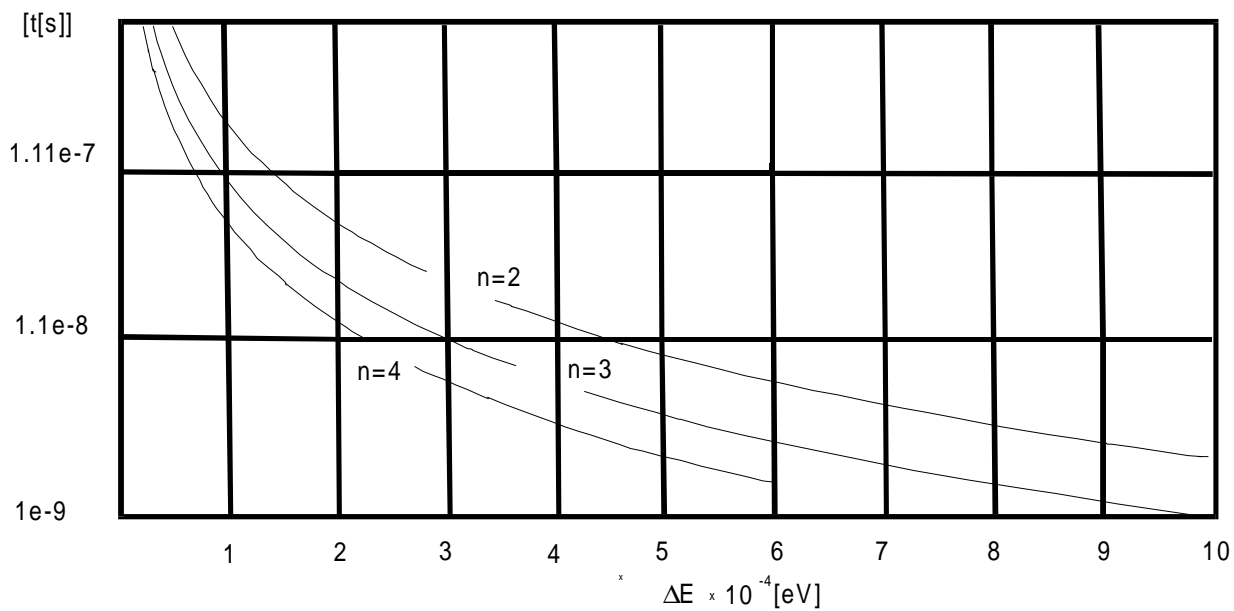

Fig.2. Time (t) versus energy ($\Delta E$) for n=2,3 and 4. These curves are symmetrical to the curves for energy differences (-$\Delta E$)

(This paper is accepted for publication in C. R. Acad. Bul. Sci. 1998)

Faculty ofPhysics  
Sofia University  
Sofia, Bulgaria